# Has Anything Changed? Tracking Long-Term Interpretational Preferences in Quantum Mechanics

Petr O. Jedlička[1], Šimon Kos[2], Martin Šmíd[3], Jiří Vomlel[4], Jan Slavík[5]

## Abstract

*As we approach the centennial anniversary of modern quantum mechanics this paper revisits the foundational debates through a new poll within the research community. Inspired by the survey by Schlosshauer, Kofler, and Zeilinger at the specialized 2011 Quantum Physics and the Nature of Reality conference, we expanded our recruitment to include a more representative sample of the broader community of physicists with the aim to reveal potential shifts in scientists' views and compare our findings with those from several previous polls. While quantum foundations still lack a consensus interpretation, our results indicate a persistent preference for the Copenhagen interpretation. This enduring support likely reflects both the educational emphasis on the Copenhagen interpretation and its pragmatic appeal in avoiding complex metaphysical questions and introducing new notions (e.g., other worlds or the pilot wave). Our findings thus underscore the relative stability of interpretational preferences over the past decades.*

[1] Faculty of Philosophy and Arts, University of West Bohemia, Sedláčkova 38, 301 00 Pilsen, Czechia; Institute of Philosophy, Czech Academy of Sciences, Jilská 361/1, 110 00 Prague 1, Czechia 0000-0002-6635-0359, jedlicka@flu.cas.cz;
[2] Department of Physics and NTIS – European Centre of Excellence, University of West Bohemia in Pilsen, Univerzitní 8, 301 00 Pilsen, Czech Republic 0000-0003-1657-9793
[3] Institute of Information Theory and Automation, Czech Academy of Sciences, Pod Vodárenskou věží 1143/4, 182 00, Prague, Czechia. 0000-0003-1140-3510
[4] Institute of Information Theory and Automation, Czech Academy of Sciences, Pod Vodárenskou věží 1143/4, 182 00, Prague, Czechia. 0000-0001-5810-4038
[5] Department of Physics and NTIS – European Centre of Excellence, University of West Bohemia in Pilsen, Univerzitní 8, 301 00 Pilsen, Czech Republic

# 1. Introduction

There is a long history of foundational debates in quantum mechanics dating back to its early years, attributable to its profound impact on established views of reality, causality, and measurement. The so-called Copenhagen interpretation, primarily based on the views of Bohr and Heisenberg, emerged in the late 1920s and early 1930s. However, the term itself was not coined until 1955 when Heisenberg (1958) introduced it in defense against the unorthodox de Broglie-Bohm alternative[6]. Notably, even Bohr and Heisenberg ostensibly differed on several issues, such as mathematical formalism and the role of measurement (Camilleri & Schlosshauer, 2015). This interpretation was initially challenged by Einstein's thought experiments, mostly on the grounds of purported incompleteness at two Solvay conferences. His line of argumentation became the core of the Einstein-Podolsky-Rosen paper in 1935, which highlighted what they perceived as existing problems of quantum mechanics but did not lead to any alternative interpretation. Equally important for subsequent developments was the notion of a hidden variable and the pilot wave theory, proposed by de Broglie and presented in his 1924 thesis and at the 1927 Solvay conference. Among other initial skeptics with lingering doubts of the Copenhagen interpretation was Schrödinger. Nevertheless, the Copenhagen interpretation acquired a status close to hegemony over the years.

Yet, its tenets have been questioned over the years by alternative theories or interpretations, among them Bohmian mechanics (1952), Everett's many-worlds interpretation (1957), and various quantum-information interpretations and their offshoots, such as the ensemble interpretation, consistent histories interpretation, and quantum Bayesianism. Interest in foundational questions has been sustained by ongoing experimental and theoretical progress, including work on Bell's inequalities and their tests, decoherence, and quantum information.

In this debate, various interpretations experience swings in popularity in the community of physicists, without a final resolution or consensus that goes beyond the accepted mathematical formalism and undisputed experimental results. On the other hand, there are also minimalist views which deny the necessity of discussing interpretations at all or consider the mathematical formalism itself to be an interpretation (Fuchs & Peres, 2000).

---

[6] Although de Broglie's and Bohm's pilot-wave interpretations are today often merged into one, they also differ in several aspects. De Broglie (1987) originally conceived the wave as a physical wave in three-dimensional space, and later further developed this theory. Bohm completed the pilot-wave theory to include the quantum potential and generalized it for the many-particle cases; however, his version does not require the physical existence of a particle-generated quantum wave. See also Croca et al (2021), or Castro, Bush, and Croca (2024).

## 2. Past polls

Scientific surveys cannot settle foundational questions in physics, but they can provide insight into how the research community's perspectives evolve in response to theoretical advancements and experimental developments. In the context of quantum mechanics, polls tracking interpretational preferences offer a snapshot of prevailing attitudes. Over the years, multiple surveys have aimed to map these preferences, examining not only the most favored interpretations but also the underlying associations between different conceptual positions.

The polls on quantum foundations in the past have been conducted mostly at conferences – by Tegmark (1998), Schlosshauer, Kofler, and Zeilinger (2013), Norsen and Nelson (2013) – or as online polls by Sivasundaram and Nielsen (2016), with dual goals: determining preferred interpretations and elucidating views on central questions (randomness, entanglement, measurement, observer, etc.), and examining whether there are any correlations between the answers. In the following, we will provide a chronologically ordered short overview of their main findings from these polls.

Tegmark's original paper (1998) sought to correct common confusion around certain postulates in Everett's many worlds interpretation and explain its metaphysical and epistemological consequences. Another reason for conducting the poll was to dispel the then widespread notion of the hegemony of the Copenhagen interpretation and the argument that MWI is only a minority stance held by physicists with "non-standard views" about science. From 48 participants in Tegmark's poll, which took place in August 1997 at the University of Maryland quantum mechanics workshop, the Copenhagen interpretation secured first place when 27% of participants voted for it. However, MWI came in second with 17%, followed by Bohmian interpretation at 8%, leading Tegmark to view these results as a sign of waning interest in the once-dominant Copenhagen interpretation. He also noted that, apart from the universal computational practices, the choice of interpretation of quantum phenomena remains purely a "matter of taste".

At the conference Quantum Physics and the Nature of Reality, held in July 2011 at the International Academy Traunkirchen in Austria, Schlosshauer, Kofler, and Zeilinger (2013) repeated the poll on the favorite interpretation, with a total of 33 participants (a mix of affiliations in physics, philosophy, and mathematics). To gain more comprehensive insights, they expanded its scope to 16 multiple-choice questions covering the most relevant issues and open questions about quantum foundations, i.e., quantum ontology and behavior, the nature of measurement and observer, relations between classical and quantum worlds, and our possible knowledge of them, as well as meta-questions concerning future developments of the field, changes in interpretations, etc. Among the findings that received strong support were the following: superpositions of macroscopically distinct states are in principle possible (67%); randomness is a fundamental concept in nature (64%); Einstein's view of quantum theory is wrong (64%); personal philosophical prejudice plays a large role in the choice of interpretation (58%); the observer plays a fundamental role in the application of the formalism but plays no distinguished physical role (55%); physical objects have their properties well defined prior to and independent of measurement in some cases (52%); the message of the observed violations of Bell's inequalities is that unperformed measurements have no results (52%).

When the answers were correlated, the authors found diverse relationships, some of which "transcended the traditional lines" such as Bohr-versus-Einstein, nonrealist-versus-realist, and epistemic-versus-ontic. While some of the patterns in data were expected, such as those who regarded the measurement problem as a pseudo-problem also tended to favor the Copenhagen interpretation (with its nuanced original arguments concerning these issues), refute Einstein's view, and consider quantum randomness as fundamental etc., others defied easy interpretation. However, the overall conclusion was that despite the mature mathematical formalism and predictive success, quantum theory still does not lend itself to a consensual interpretation.

In 2013, Norsen and Nelson (2013) ran a sequel to this poll with 76 participants at a conference Quantum Theory Without Observers III, held in Bielefeld, Germany. Here, the results again demonstrated the existence of "sub-communities with quite different views," and led the authors to believe that "there is probably even significantly more controversy about several fundamental issues" than had been revealed in previous polls. The authors also found strong support for the de Broglie–Bohm interpretation among the participants (63%), compared to only 4% for the Copenhagen interpretation, which they inferred was the result of the invitation process favoring scientists with kindred "realist" views.

Sivasundaram and Nielsen (2016) conducted an online poll with a slightly different set of questions, polling 149 physicists of all specializations from eight universities in several countries, which secured a sample of physicists that is the most representative of all polls. The Copenhagen interpretation garnered the highest rate with 39%, with only 6% for Everett's interpretation (many worlds or/and many minds), 2% for de Broglie–Bohm interpretation, and 17% in total for other interpretations. Altogether, 36% of the sample had no preferred interpretation. Crucially, the authors came to the conclusion that "foundational concepts in quantum mechanics are still a topic that only a minority of physicists are familiar with," although they deem it important.

## 3. Methodology

### 3.1. Questionnaire

Our questionnaire was based on the 16 multiple-choice questions formulated by Schlosshauer, Kofler, and Zeilinger. For the polls, we selected 10 questions that we deemed most relevant (there was a limited total length for the poll questionnaire). Prior to full deployment the questionnaire was piloted with several physicists not specializing in quantum foundations. Feedback led to minor wording adjustments that improved clarity without changing substantive content. No further validation was performed.

Here are the questions and response options (for those used in further analysis with their italicized abbreviations):

Q1: What is your opinion about the randomness of individual quantum events (such as the decay of a radioactive atom)? (*Rnd*):
The randomness is only apparent (*RndApp*); There is a hidden determinism (*HidDet*); The randomness is irreducible (*RndIrred*); Randomness is a fundamental concept in nature (*FundRnd*).
Q2: Do you believe that physical objects have their properties well defined prior to and independent of measurement? (*Wdp*):
Yes, in all cases: (*Yes*); Yes, in some cases: (*Some*); No; I'm undecided (*NotKnow*).

Q3: Is the quantum mechanical description of reality by a wave function complete? To which alternative are you inclined? (*Wdr*):

No, the wave function description of reality is not complete (A. Einstein) (*Einstein*); The description of reality by a wave function is complete (N. Bohr) (*Bohr*); I don't know, there is another way. (*Other*)

Q4: The measurement problem is (*Msr*):

A pseudo-problem (*Pseudo*); Solved by decoherence (*Decoher*); Solved/will be solved in another way; A serious difficulty threatening quantum mechanics (*Serious*); None of the above (*None*).

Q5: What interpretation of quantum states do you prefer? (Qs):

Epistemic/informational (*Epistem*); Ontic (*Ontolog*); A mix of epistemic and ontic (*Comb*); Purely statistical (e.g., ensemble interpretation) (*Stat*); Other.

Q6 The observer (*Obs*):

Is a complex (quantum) system: (*Complex*); Should play no fundamental role whatsoever; Pays a fundamental role in the application of the formalism but plays no distinguished physical role: (*ApplForm*); Plays a distinguished physical role (e.g., wave-function collapse by consciousness).

Q7. Which interpretation of quantum mechanics do I prefer? (*Qm*)

Copenhagen (quantum mechanics is a complete description of the microworld, measurement is an intervention from the classical world) (*Copenhag*); Everett's (in measurement the world breaks down into multiple worlds – all possibilities are realized in some world – i.e. many worlds or many minds) (*Everett*); De Broglie-Bohm (particle moves in a deterministic potential and randomness is in the initial conditions)[7] (*Bohm*); Other (*Other*).

Q8. Superpositions of macroscopically distinct states (*Val*): Are in principle possible (*PrincPos*); Will eventually be realized experimentally (*Experim*); Are in principle impossible (*ImpPrinc*); Are impossible due to a collapse theory.

Q9. How often have you switched to a different interpretation? (*Chn*):

Never (*Never*); Once (*Once*); Several times (*Several*); I have no preferred interpretation: (*NoPref*).

Q10. How much is the choice of interpretation a matter of personal philosophical prejudice? (*Phl*):

---

[7] Although de Broglie's interpretation is often labelled deterministic, it should, *sensu stricto*, be regarded as merely causal (de Broglie 2021).

A lot (*Much*); A little (*Little*); Not at all (*NotAtAll*).

## 3.2. Data collection

The poll was conducted as part of a project on scientific objectivity at the four largest Czech research institutions: Czech Academy of Sciences, Charles University in Prague, Masaryk University in Brno, and University of South Bohemia, along with their research institutes. It was hosted on a local online platform and administered via email, which provided main information about the project. Respondents were incentivized with a coupon for an online bookstore upon completion.

To maintain consistency with previous polls, we included in our analysis only those responses from participants whose main specialization was in mathematics, physics (including applied physics), and information science. We imposed a cut-off time limit and excluded entries completed in under eight minutes, as we considered this duration to be the minimum to meaningfully respond to the entire questionnaire. This approach yielded a total of 40 valid questionnaires, although the response rate varied per question since answering the questions was not mandatory (the number of respondents N is specified for each question). Additionally, the response choices were limited to only one option in our poll, so we make only relative comparisons with the results from other polls in our paper.

## 3.3. Statistical Analysis

In the results section, we first provide simple descriptive statistics comprising of the distributions of response options in the graphical representation, which corresponds to the original paper by Schlosshauer, Kofler, and Zeilinger. These results were subsequently analyzed and compared with the Schlosshauer, Kofler, and Zeilinger's results and results of the other polls. In the second part, we opted for Bayesian network analysis (BNA) and Chi-square tests combined with the Spearman correlations to reveal further relationships between the variables.

To investigate the interrelationships between questions, we employed Bayesian networks (BNs), which are probabilistic models that utilize graphs to visualize independence and dependence relations between variables (Pearl, 1988; Jensen & Nielsen 2007; Koller & Friedman 2009). These models are particularly well-suited for nominal variables, which is the case for our questions. To ascertain the graphical structure of the BN model, we employed a learning algorithm that maximizes the Akaike Information Criterion (AIC), which is calculated as the log-likelihood of observed data given the model minus the number of model parameters. The penalty component of AIC serves to prevent overfitting of data by overly complex models. All computations were performed in R using the bnlearn package (R Core Team 2021).

The resulting model is presented as a Complete Partially Directed Acyclic Graph (CPDAG) of the learned Bayesian network, a hybrid graph that can contain both directed and undirected edges that represents an equivalence class of Bayesian networks. (The CPDAG has the same skeleton as all BNs from the equivalence class it represents. Furthermore, only edges that have the same direction in all Bayesian networks of the equivalence class are directed in concordance with these BNs. See R Core Team (2021) for details.) To assess the degree of confidence in each edge, we ran the learning algorithm 40 times, removing one respondent in each iteration, and computed the relative frequency of each edge's presence in the model.

Additionally, we investigated the relationship between variables by examining correlations of binary indicators for selected response options. To quantify the dependence, we used the Chi-square test, and to determine the direction of the relation, we employed the Spearman correlation. To obtain robust results, we proceeded in a similar way as above: we computed the correlations 40 times, each time excluding a single observation. Consequently, we calculated the relative frequencies of test results that were significant at the 0.05 level.

The code and the dataset are available at https://gitlab.cesnet.cz/utia/public/quantum-interpretations .

## 4. Results and discussion

In this section, we summarize the results of our poll, providing a short commentary that also discusses and compares them with the results of previous polls.

### 4.1. Descriptive statistics

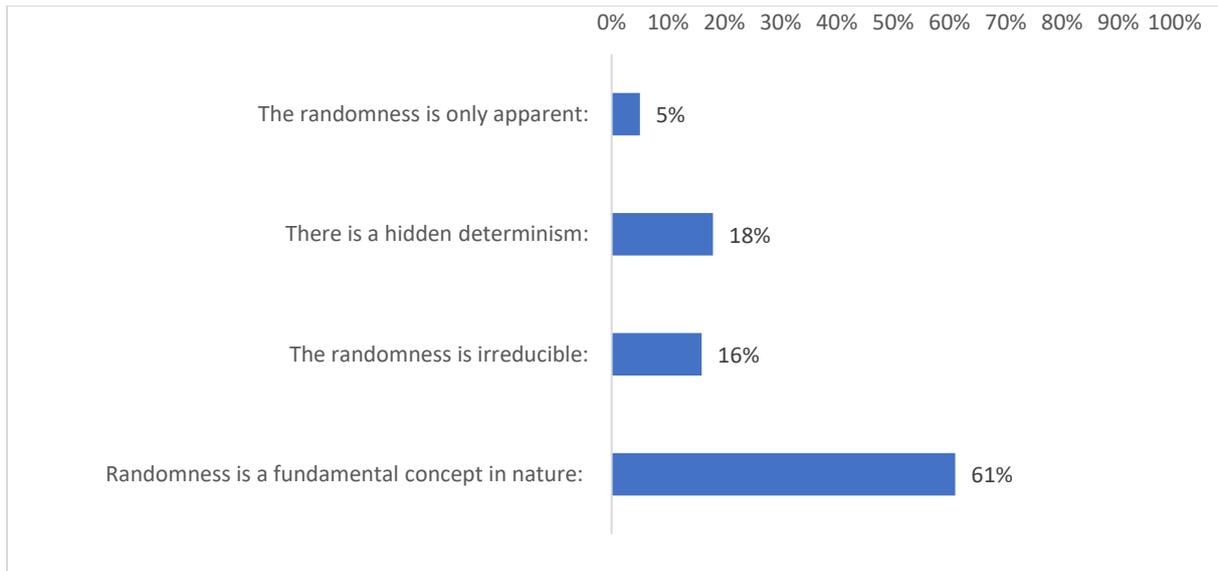

**FIG. 1. (Q1, RnD): What is your opinion about the randomness of individual quantum events (such as the decay of a radioactive atom)? (N= 38)**

A majority of the participants in our poll supported the idea that randomness is a fundamental concept in nature, with over three-fifths choosing this answer (Figure 1). This view is prevalent in the community of physicists, as similar results were also obtained in the Schlosshauer, Kofler, and Zeilinger and Sivasundaram and Nielsen polls. The exception was the Norsen and Nelson poll, where it received little support (less than a quarter). Conversely, there was low support for the ideas that randomness is only apparent or that there is a hidden determinism, again with the notable exception of the Norsen and Nelson poll.

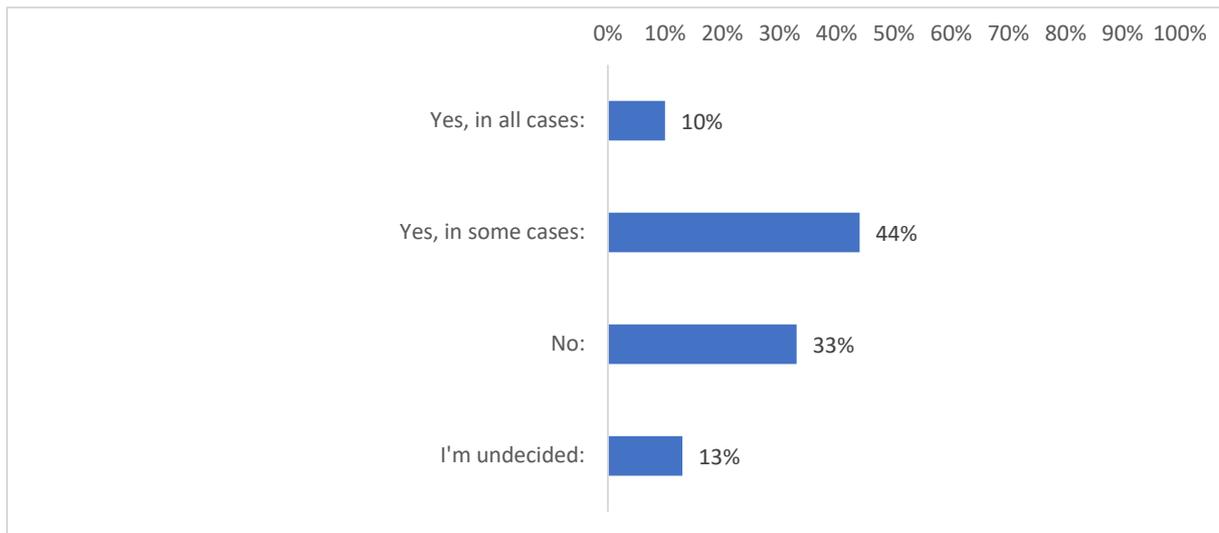

**FIG. 2. (Q2, Wdp): Do you believe that physical objects have their properties well defined prior to and independent of measurement? (N=39)**

The most common response in our poll was 'yes, in some cases' (Figure 2), which was consistent with the Schlosshauer, Kofler, and Zeilinger and Norsen and Nelson polls. In the Norsen and Nelson poll, participants generally believed that the properties are well-defined in all cases, or at least in some cases, whereas Sivasundaram's and Nielsen's results diverged substantially, with 'no' being the most favored answer.

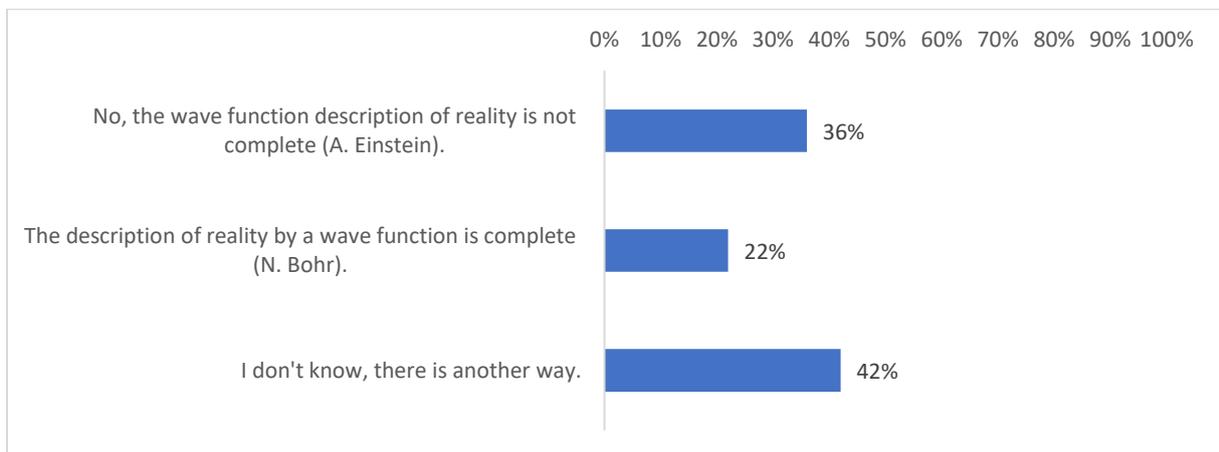

**FIG. 3. (Q3, Wdr): Is the quantum mechanical description of reality by a wave function complete? To which alternative are you inclined? (N=36)**

Into this single question, we condensed the essence of questions 3 and 4 from the Schlosshauer, Kofler, and Zeilinger questionnaire, focusing on Einstein's view of quantum mechanics and Bohr's view of quantum mechanics. Our results indicated somewhat stronger leanings towards the notion that the wave function description of reality is incomplete, aligning with Einstein's stance and contrary to Bohr's views (Figure 3). However, over two-fifths of the scientists were undecided and expected another solution. Previous polls showed divergent responses: participants in the Schlosshauer, Kofler, and Zeilinger poll largely considered Einstein's views to be incorrect or likely to be disproven in the future (three quarters); Norsen and Nelson poll participants generally disliked both Bohr's views (close to three quarters), as well as Einstein's (almost half).

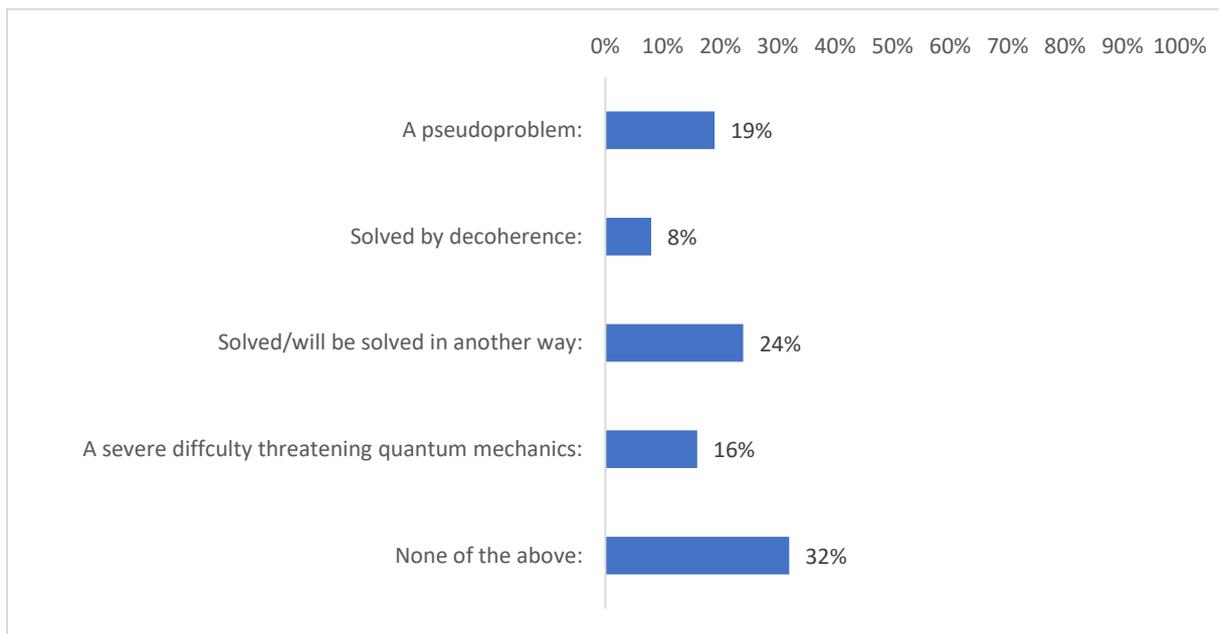

**FIG. 4. (Q4, Msr): The measurement problem is: (N=37)**

A third of the participants in our poll remained undecided (Figure 4). Other polls did not show a general consensus either. For instance, in the Sivasundaram and Nielsen poll, a third of the participants did not prefer any of the variants, whereas in both the Schlosshauer, Kofler, and Zeilinger and Norsen and Nelson polls, the most preferred (although vaguely defined) solution was that the problem will be "solved in another way."

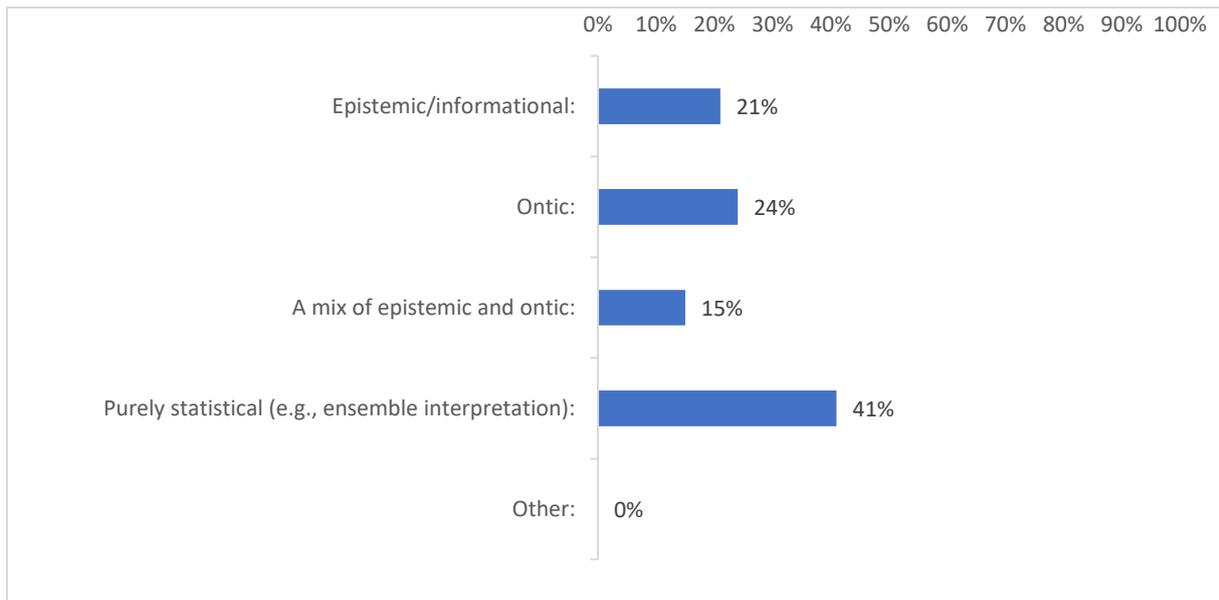

**FIG. 5. (Q5, Qs): What interpretation of quantum states do I prefer? (N=34)**

In our poll, the "purely statistical (e.g., ensemble interpretation)" prevailed (Figure 5). The results in other polls are mixed, with Norsen and Nelson participants showing a higher preference for an ontic interpretation and Schlosshauer, Kofler, and Zeilinger favoring a mix of epistemic and ontic interpretations.

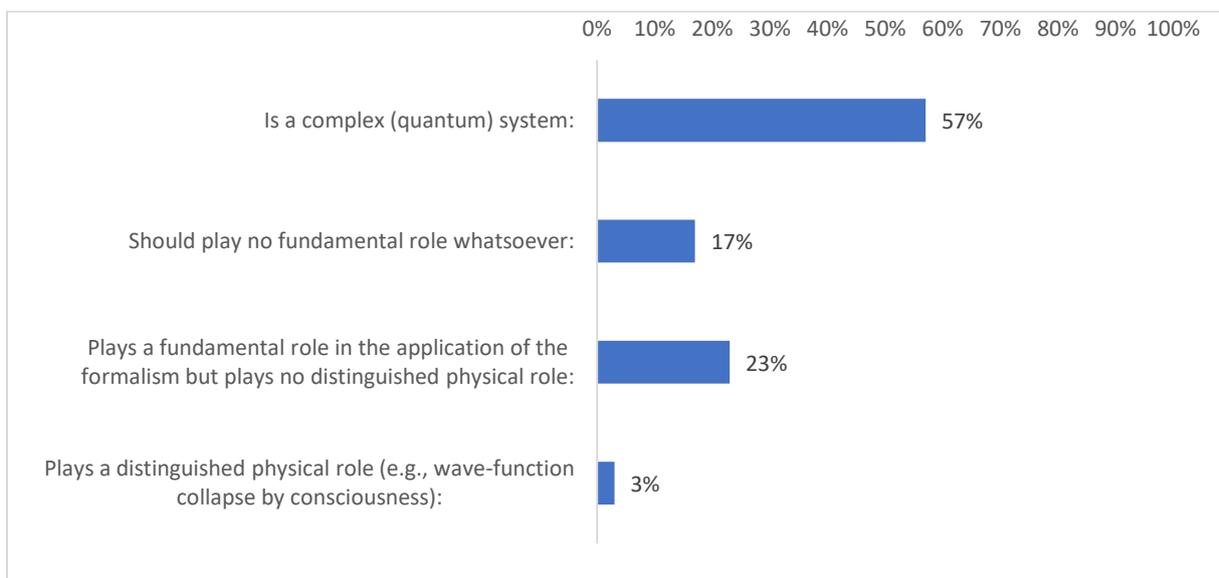

**FIG. 6. (Q6, Obs): The observer (N=35)**

The dominant view in our poll is that the observer is a complex (quantum) system (Figure 6), which was also the most common view in the Sivasundaram and Nielsen poll. Results in the other two polls (Schlosshauer, Kofler, and Zeilinger and Norsen and Nelson) are inconclusive. Only a small, single-digit fraction in our poll, as well as in Schlosshauer, Kofler, and Zeilinger and Norsen and Nelson polls, subscribe to the view that the observer plays a distinguished physical role (e.g., wave-function collapse by consciousness).

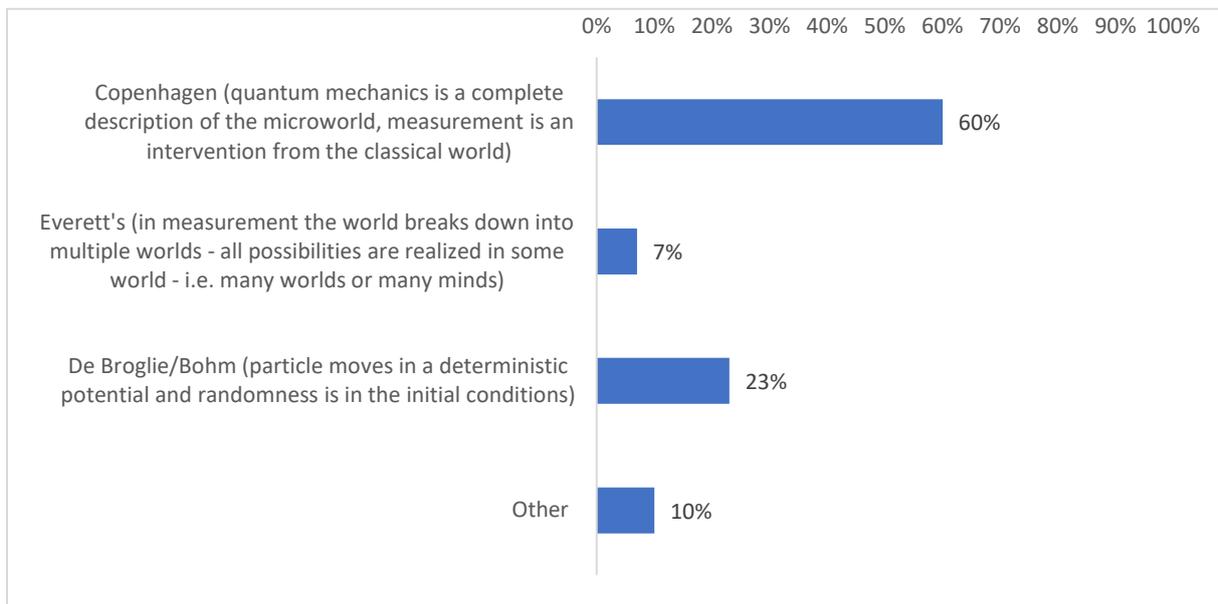

**FIG. 7. (Q7, Qm): Which interpretation of quantum mechanics do I prefer? (N=30)**

Following pilot testing, we simplified the options for this question to include the three most popular interpretations (Copenhagen, Everett's, de Broglie-Bohm's) but also left room for participants to propose their preferred interpretation ("Other" option) in an open question, among which, for example, the "Joos-Zeh" variant occurred. In our poll, the Copenhagen interpretation received the majority of votes and was the most favored across the board (Figure 7), as it also garnered the highest support in the Schlosshauer, Kofler, and Zeilinger and Sivasundaram and Nielsen polls. The noteworthy exception was the Norsen and Nelson poll, in which de Broglie-Bohm's theory prevailed, and Copenhagen lost significantly – according to the authors due to the biased attendance at the event.

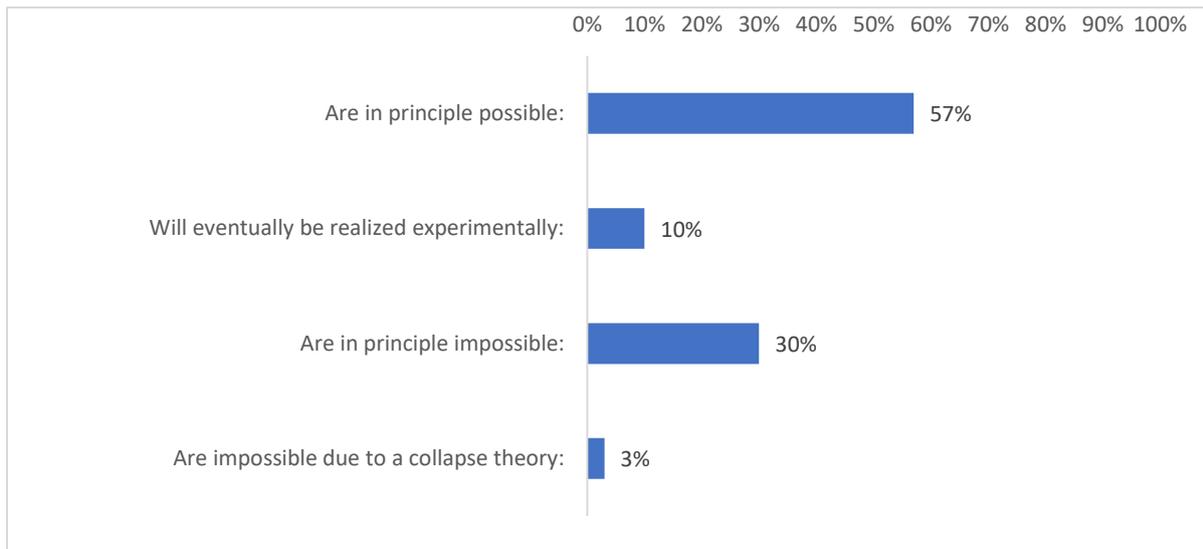

**FIG. 8. (Q8, Val): Superpositions of macroscopically distinct states. (N=30)**

This question produced universally consistent answers across all four polls (Figure 8), with the majority of participants supporting the idea that superpositions of macroscopically distinct states are in principle possible (over half in all polls), contrasted with the lowest support for the view that superpositions are impossible due to a collapse theory (in single digits in all polls).

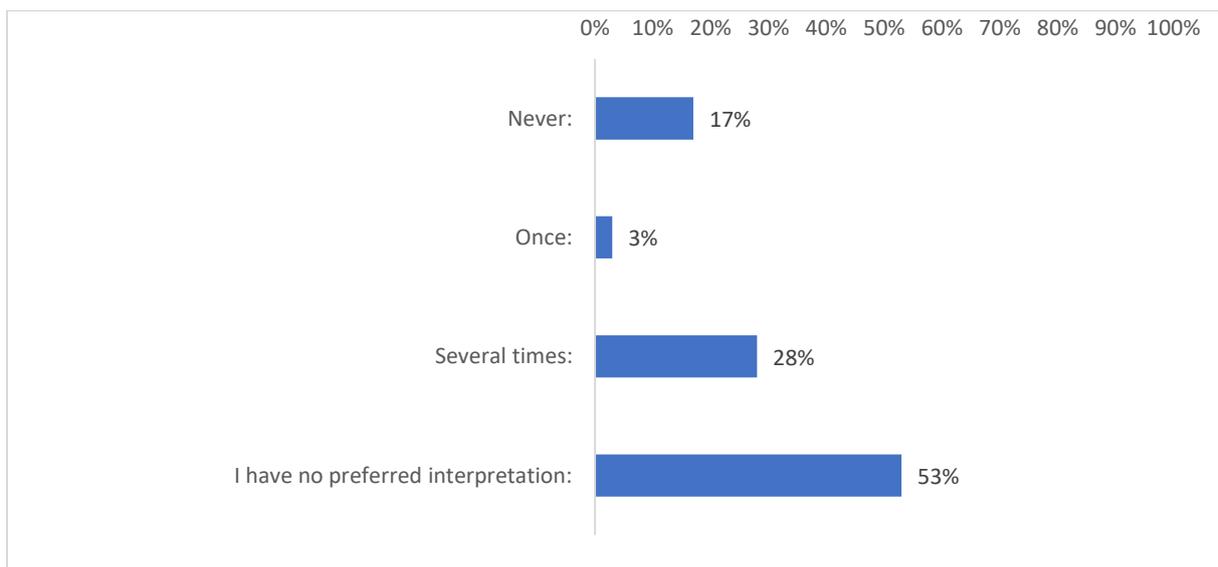

**FIG. 9. (Q9, Chn): How often have you switched to a different interpretation? (N=36)**

In our poll, the single most frequent answer to this meta-question was "no preferred interpretation", and the least frequent was "once" (Figure 9). In contrast, in all three other polls (Schlosshauer, Kofler, and Zeilinger; Norsen and Nelson; and Sivasundaram and Nielsen), the most popular choice was that participants had never switched to a different interpretation.

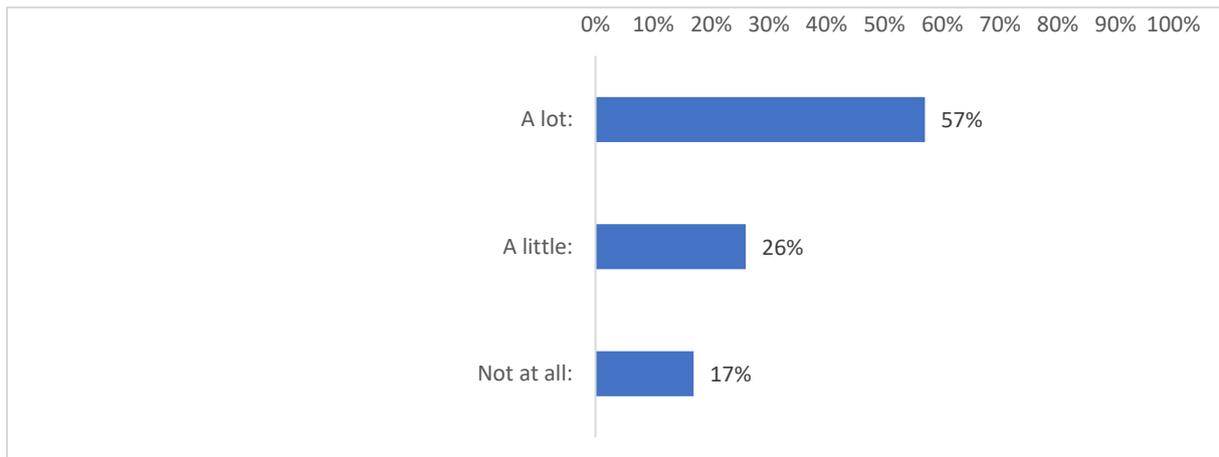

**FIG. 10. (Q10, Phl): How much is the choice of interpretation a matter of personal philosophical prejudice? (N=35)**

For this meta-question (Figure 10), there was universal agreement in all polls, except for Sivasundaram's and Nielsen's (where this question was not posed), about the claim that philosophical prejudice plays a significant role in the choice of interpretation, indicating a high level of self-awareness about the vast influence of factors outside quantum mechanics itself.

### 4.2. Bayesian network analysis

For a deeper analysis, we employed the Bayesian network Analysis (BNA). The resulting Bayesian networks model is presented in Figure 11 in the form of its CPDAG. It is noteworthy that our BN model contains exclusively undirected edges.

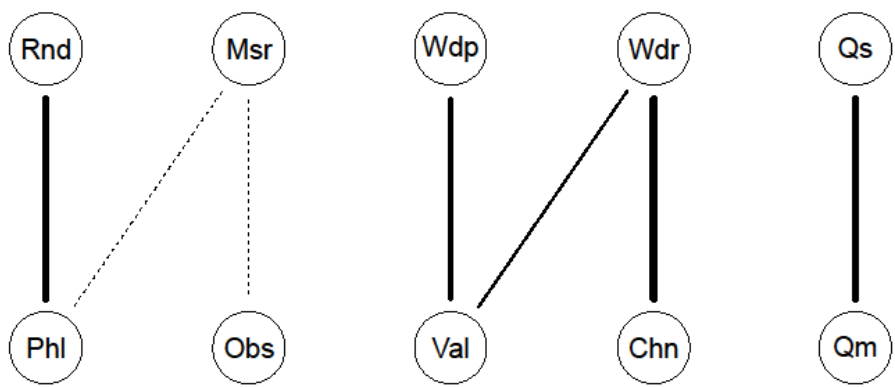

**FIG. 11. A Bayesian network model presented in the form of its CPDAG. In the graph, the width and the type of an edge indicate the degree of confidence in that edge.**

Subsequently, in the resulting model, three edges were identified as consistently present across all models learned by the method discussed in Section 3.3, thus showing the dependence of variables Rnd (Q1) and Phl (Q10), variables Wdr (Q3) and Chn (Q9), and variables Qs (Q5) and Qm (Q7), all with relative frequencies 1.0.

The remaining four edges exhibited the following relative frequencies: The relative frequencies of the edges Wdp-Val (Q2 and Q8), Wdr-Val (Q3 and Q8), Msr-Phl (Q4 and Q10), and Msr-Obs (Q4 and Q6) were 0.925, 0.875, 0.425, and 0.250, respectively. Edges with a relative frequency lower than 0.25 were excluded from the model.

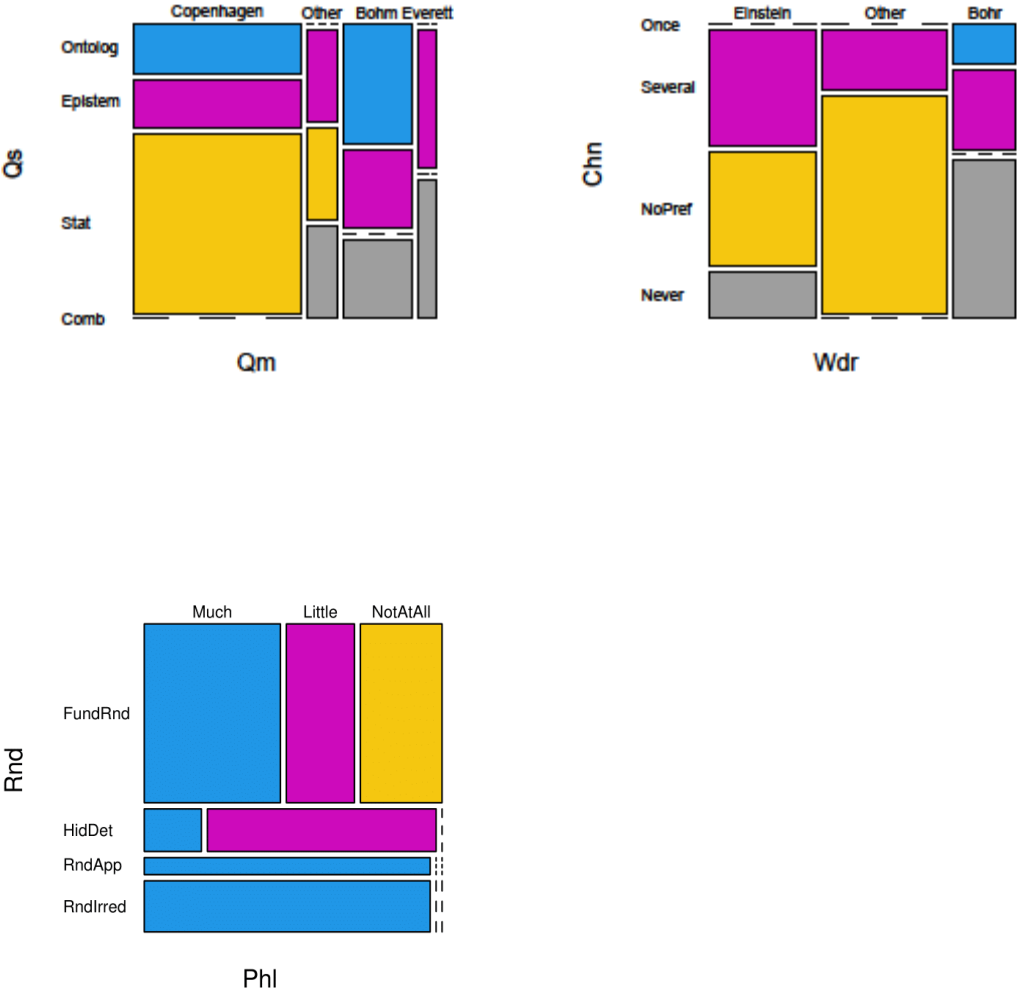

**FIG. 12. The mosaic plot presents probability distributions corresponding to the three most significant edges. These plots illustrate the dependency patterns between answers to the two corresponding questions.**

In the first plot of Figure 12, we can see that the Copenhagen interpretation is predominantly, though not entirely, associated with a purely statistical description, in contrast to Everett's, which is mostly associated with epistemic, or combined epistemic and ontic view of the quantum states, and de Broglie-Bohm's interpretation, which is mostly associated with the ontic view (and less with epistemic and combined view), which is of little surprise.

In the second plot of Figure 12, the association between the preferred views on completeness of the quantum mechanical description of reality (Wdr) and the changes in interpretations (Chn) suggests that those subscribing to Bohr's views are not likely to change their interpretation and the changes are more common for those who subscribe to Einstein views or prefer other views. Similarly, Sivasundaram and Nielsen found a strong correlation between those preferring the Copenhagen interpretation and those who never changed their interpretation.

The responses also showed association between "other views" on the completeness and absence of any preferred interpretation ("No Preference"). This finding is consistent with the finding in Schlosshauer, Kofler, and Zeilinger paper, where this relationship had a "medium correlation".

The third plot of Figure 12 indicates that the personal philosophy is somewhat involved in the choice of interpretation that is shared across the board, but predominately by those who think that randomness is irreducible or that it is "only apparent" (RndApp).

### 4.3. Chi-square tests and correlation analysis

Additionally, we examined the relationships between individual values in the responses using Chi-square test and Spearman correlation. The comprehensive results can be seen in Figure 13. (There has never been a case where a significant relationship between two variables changed direction from one run to the next.)

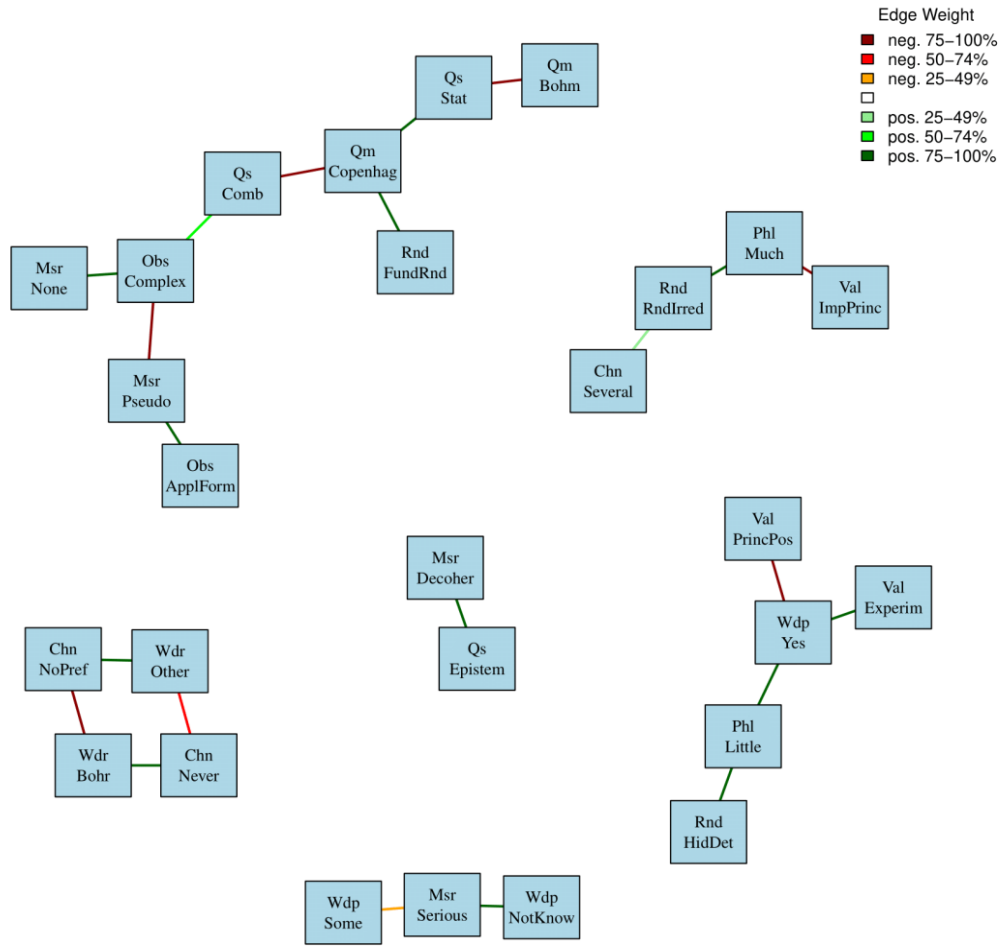

**FIG. 13.** The Chi-square independence test plot, where green edges indicate positive correlations between variables, and red or yellow edges signify negative correlations. The intensity of the edge's shade represents the relative frequency of significant test results at the 0.05 level across 40 subanalyses. In each subanalysis, the independence test was performed on the full sample, excluding one subject at a time.

Several relationships between the responses stand out here. For example, there is a positive relationship between those who believe physical objects have well-defined properties before measurement (Wdp Yes) and that philosophical bias plays little role (Phl Little) suggesting a connection between realist interpretations and a dismissal of philosophical influence. Respondents who tend to believe that philosophical bias has little influence (Phl Little) also support hidden determinism (Rnd HidDet). We can assume the underlying factor is that both groups seek objective explanations independent of personal bias.

In the cluster on the right, respondents who believe the observer plays a role in applying formalism but not a fundamental physical role (Obs ApplForm) tend to see the measurement problem as a pseudo-problem (Msr Pseudo). On the contrary, the belief that the measurement problem is a pseudo-problem is negatively correlated with the belief that the observer is a complex quantum system (Obs Complex). And those who consider the observer a complex quantum system also do not often have a definitive view on the measurement problem (Msr None). In Schlosshauer, Kofler, and Zeilinger's paper, we find analogous relationships – e.g. those who dismiss measurement problem as a pseudo-problem also do not ascribe a distinguished physical role to the observer etc.

In the same cluster, we also found a strong positive relationship between those who prefer the Copenhagen interpretation (Qm Copenhag) and those who view the randomness fundamental and favor a statistical interpretation of quantum states (Qs Stat) reflecting the traditional link (although not identity) between Copenhagen interpretation and a statistical/ensemble view of quantum mechanics. On the other hand, statistical interpretation of quantum states (Qs Stat) is negatively correlated with the de Broglie-Bohm interpretation (Qm Bohm). Those who prefer a combination of epistemic and ontic interpretation of quantum states (Qs Comb) weakly tend to consider the observer as a complex quantum system (Obs Complex) and are less likely to adopt the Copenhagen interpretation (Qm Copenhag).

In the remaining clusters, we can see that those who adopt an epistemic interpretation of quantum states (Qs Epistem) also believe that the measurement problem will be solved by decoherence (Msr Decoher). The correlation analysis also showed a positive (and expected) link between Bohr's views on the completeness of the quantum mechanical description of reality (Wdr Bohr) and no change in opinions (Chn Never) and a positive link between those who do not subscribe to either Einstein's or Bohr's views (Wdr Other) and those who do not have a preferred interpretation (Chn NoPref).

## 5. Conclusion

Our results affirm the strong position of the Copenhagen interpretation (Q7), which is as we found typically often in respondents accompanied with the corresponding views: that randomness is fundamental (Q1), and prefer statistical interpretation of quantum states (Q5). Copenhagen interpretation is also negatively correlated with the view that the observer is a complex quantum system (Q6) which could be explainable by its operational viewpoint of the observer. Similarly, those who ascribe only a formal role to the observer tend to see the measurement problem as a pseudo-problem (Q4). Competing interpretations garnered opposing views, for instance, the de Broglie-Bohm's interpretation was negatively correlated with statistical interpretation of quantum states (Q5).

Another significant finding was that those who hold Bohr's views (Q3) usually haven't experienced any changes in interpretations during their careers. This could be explained by the fact that Copenhagen interpretation is the default interpretation taught in undergraduate and graduate courses. Consequently, for many physicists who do not specialize in quantum foundations, it very likely remains their preferred interpretation. Its popularity may also be elevated because it seemingly abstracts from the most vexing metaphysical and epistemological questions.

Altogether, our results reveal a continuing deep rift within the physicists' community as the responses in our poll tend to be distributed along the whole spectrum, although some similar patterns did occasionally emerge in our and the previous polls, such as in views about randomness (Q1), measurement (Q4), and superpositions (Q8). This situation is atypical for other canonical physical theories, although historical parallels of long-term disagreements of various kinds certainly exist. In our opinion, this situation is not likely to be amended anytime soon until a significant theoretical or experimental progress is achieved in this or another field.

It is, however, obvious that in previous polls, the method of data collection played a crucial role. These polls were highly informal and, therefore, prone to sampling biases originating both in the invitation process and in the actual poll participation, which influenced their results. These biases also inadvertently impacted our poll; however, our poll together with the Sivasundaram and Nielsen poll, were likely the most representative ones, as the data were not collected from a subgroup involved in quantum foundations or from a group specialized in a particular interpretation.

Despite these reservations, our findings confirm steady and relatively strong support for the Copenhagen interpretation among those who have picked an interpretation to begin with, which has been quite consistent pattern over a 25-year period but probably also earlier when such polls were not common. Thus, the fluctuations regarding interpretations and some seminal questions appear to stem mostly from the biased audience at specific quantum foundation events rather than from the evolution over the time. Our findings, therefore, do not lend much support to Tegmark's initial claim from 1997 that "the prevailing view on the interpretation of quantum mechanics appears to be gradually changing," as the preferences in interpretations have remained rather stable since then.

Most of all, however, our results suggest that there is still a significant amount of uncertainty in the community, and many practitioners simply remain undecided about their preferences – particularly the general physicists' community approached in our and the Sivasundaram and Nielsen poll, which in greater numbers stated that they do not have a preferred interpretation.

Physicists, for the most part, seem to be aware of the inherent "shakiness" of quantum foundations, as is apparent from the answers to the meta-questions concerning the evolution of one's own views and about the influence of personal philosophical prejudice (Q9 and Q10). Here, the responses indicate that physicists understand that the lack of a more comprehensive theory leaves substantial room for their personal views or prejudices, as the most common answer in all polls was that the prejudice "matters a lot" in the choice of interpretation.

## Statements and Declarations

## Funding

P.J. was supported by Grant No. 18-08239S of the Czech Science Foundation. Š.K. was supported by the project Quantum materials for applications in sustainable technologies (QM4ST), funded as Project No. CZ.02.01.01/00/22_008/0004572 by Programme Johannes Amos Comenius, call Excellent Research.

## Competing interests

The authors have no relevant financial or non-financial interests to disclose.

## Data availability

The data that support the findings of this study are available within the article.

## Contribution statement

Petr Jedlička: conceptualization, data curation, investigation, formal analysis, methodology, writing – original draft; Šimon Kos: conceptualization, data curation, investigation; Martin Šmíd: formal analysis, methodology, visualization; writing – original draft; Jiří Vomlel: formal analysis, methodology, visualization, writing – original draft; Jan Slavík: conceptualization, methodology.